# Galactic Dynamics and Local Dark Matter

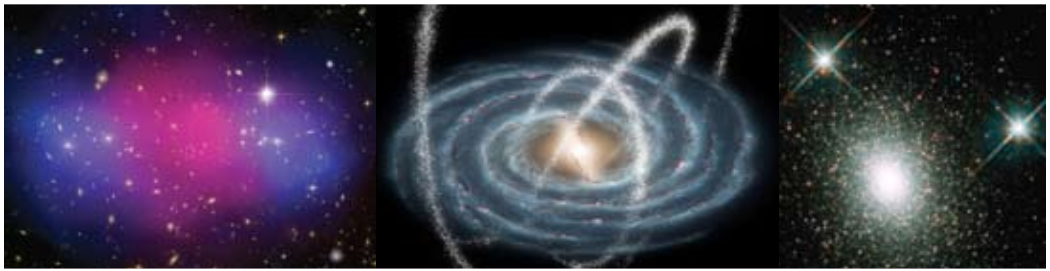

**Steven R. Majewski** (Univ. Virginia), **James Bullock** (UCI), **Andreas Burkert** (Univ.-Sternwarte München), **Brad Gibson** (Univ. of Central Lancashire), **Eva Grebel** (Univ. Basel), **Oleg Y. Gnedin** (Univ. Michigan), **Puragra Guhathakurta** (UCSC), **Amina Helmi** (Univ. Groningen), **Kathryn V. Johnston** (Columbia Univ.), **Pavel Kroupa** (Univ. Bonn), **Manuel Metz** (Univ. Bonn), **Ben Moore** (Univ. Zurich), **Richard J. Patterson** (Univ. Virginia), **Edward Shaya** (Univ. Maryland), **Louis E. Strigari** (Stanford Univ.), and **Roeland van der Marel** (STScI)

**ABSTRACT**

*The concordance Λ Cold Dark Matter (ΛCDM) model for the formation of structure in the Universe, while remarkably successful at describing observations of structure on large scales, continues to be challenged by observations on galactic scales. Fortunately, CDM models and their various proposed alternatives make a rich variety of testable predictions that make the Local Group and our own Milky Way Galaxy key laboratories for exploring dark matter (DM) in this regime. However, some of the most definitive tests of local DM require μas astrometry of faint sources, an astrometric regime that is a unique niche of SIM Lite. This chapter explores the important and distinct contributions that can be made by SIM Lite in the exploration of galaxy dynamics and DM on galaxy scales and that have cosmological consequences. Key areas of potential SIM Lite exploration include (1) measuring the shape, orientation, density law, and lumpiness of the dark halo of the Milky Way and other nearby galaxies, (2) determining the orbits of Galactic satellites, which may be representatives of late infall from the hierarchical formation of the Milky Way, (3) ascertaining*

*the distribution of angular momentum and orbital anisotropy of stars and globular clusters to the outer reaches of the Galactic halo, dynamical properties that hold clues to the early hierarchical formation of the Galaxy, (4) measuring the physical nature of DM by placing strong constraints on the phase space density in the cores of nearby dSph galaxies, and (5) reconstructing the dynamical history of the Local Group through the determination of orbits and masses of its constituent galaxies.*

## 4.1 Understanding Structure Formation on Galactic Scales

### 4.1.1 Pushing the Frontiers of Near-Field Cosmology

Empirical studies of the structure and evolution of the Universe must navigate a trade space that includes, at one extreme, surveys of galaxies to high redshift — which yield critical sampling of the luminous Universe in the time dimension but with ever-decreasing levels of detail as distances increase — and, at another extreme, exploration of galaxies in the local Universe — which can provide singularly detailed views of structure and dynamics, but with coarser time resolution. With the latter strategy, often called "near-field cosmology," we see galaxies only in the recent epoch, but they still harbor ancient stellar populations that formed at the earliest epochs of galaxy formation. The spatial, chemical and kinematical distributions of stars of different ages in a galaxy provide valuable information about the properties of the environment in which they formed and allow us archaeologically to reconstruct evolutionary histories.

Both of these strategies, far- and near-field cosmology, provide complementary constraints and contexts that successful theories of the evolving Universe must simultaneously satisfy and describe, and many recent advances in understanding the evolution of galaxies and the Universe are directly tied to great strides in the development of each approach. While technological advances have now allowed far-field cosmology to see galaxies (or at least their central quasar beacons) to $z \sim 6$, so too has the domain of near-field cosmology, most exquisitely realized through the study of resolved stars, been pushed out from the solar neighborhood to the remote reaches of the Milky Way, into the Local Group and beyond. SIM Lite will make a significant and unique contribution to expanding the frontiers of near-field cosmology across this realm by providing unprecedentedly accurate sensitivity to the motions of stars in galaxies on Mpc scales. New data on the dynamics of stars that SIM Lite can measure within the Milky Way and in galaxies in the Local Group — which we expect to be more or less typical of spiral galaxies and modest density galaxy groups at the current epoch — can provide critical leverage on still perplexing questions regarding the assembly of structure on small scales that cannot be gleaned from the simple snapshots and global dynamical measurements derived for galaxies at higher redshifts.

### 4.1.2 Testing CDM Models Within the Local Group

Since the seminal study of Searle and Zinn (1978), the notion of accretion of "subgalactic units," including "late infall," has been a central concept of Milky Way and galaxy evolution studies. Since then, large-scale *N*-body simulations of the formation of structure in the Universe in the presence of DM (and dark energy) have also shown galaxies like our Milky Way — as well as all large structures in the Universe — building up hierarchically (e.g., Figure 4-1; also see Figure 4-7). The predicted large-scale filamentary structure of the Universe is observed in the distributions of galaxies as mapped by extensive galaxy redshift surveys, while the presence of substructure within galaxies and galaxy clusters provides strong support for the merger hypothesis, so that bottom-up structure formation on all scales has become the commonly accepted paradigm.

Figure 4-1. One of the highest-resolution, fully cosmological hydrodynamical disk galaxy simulations (showing only the gas distribution) to date, and the first generated with a grid-based code (a complementary approach to particle-based codes, like that shown in Figure 4-7). This model uses the RAMSES Adaptive Mesh Refinement code to reach a spatial resolution better than 100 pc at the present day. (Model image from B. Gibson.)

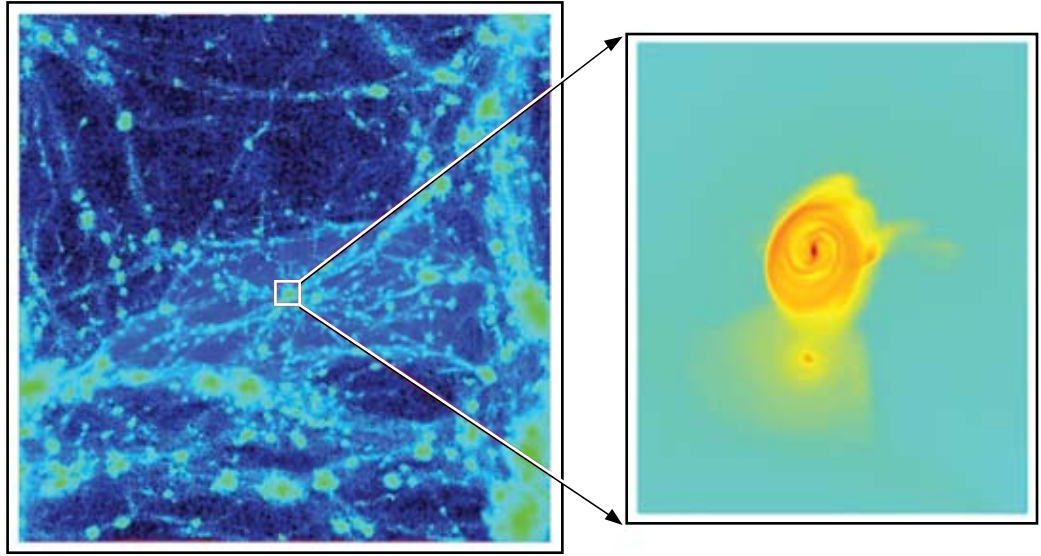

The active merging history on all size scales demonstrated by high-resolution numerical simulations has had remarkable success in matching the observed properties of the largest structures in the Universe. However, the specific paradigm including cold dark matter (CDM) is still a challenge to reconcile with the observed properties of nearby (i.e., present-day) structures on galactic scales, and this discrepancy remains one of the primary challenges to be resolved by near-field cosmology.

While Galactic halo stars, globular clusters, and satellite galaxies represent a small fraction of the baryonic mass of the Milky Way, these stellar systems are "Rosetta Stones" to understanding the origin of our galaxy and the nature of its DM halo. Recorded in the orbital dynamics of halo stars and globular clusters are unique signatures of early galactic formation processes, while the bulk motions of satellite galaxies hold information about the infall processes of more recent galaxy assembly. The kinematics of these systems also provide sensitive probes of the current DM distribution throughout the galaxy.

In this context, aspects of various well-known discrepancies between CDM theory and observations, such as the "missing satellites problem" (§4.2.1, §4.4.1), the "central cusps problem" (§4.5.1), and problems with the angular momentum distributions in galaxies (§4.4.1) are ripe for exploration in our local universe. The Local Group — and in particular the Milky Way and its satellite system — have become particularly important laboratories for testing specific predictions of the CDM models and their alternatives. Most especially, high-accuracy µas astrometric observations enabled by SIM Lite will allow definitive tests of dynamical effects specifically predicted by these models and allow a means to determine the present spatial distribution and primordial phase space distribution of DM in the Local Group. While some tests of local DM do fall within the "Gaia-sphere" — that region of the Milky Way accessible to the European Space Agency's Gaia mission — several key experiments can only be carried out definitively within a part of the (apparent magnitude)-(astrometric precision) parameter space that is out of reach of Gaia and uniquely accessible to SIM Lite's interferometric capability.

While current cosmological simulations have not yet succeeded in spawning structures that truly resemble real galaxies (e.g., Abadi et al. 2003), the present brisk pace of development of these models (e.g., as shown in Figure 4-1) can be expected to continue with eventual convergence in the coming decade to significantly improved and detailed predictions for the formation of the Local Group, the Galaxy, and its stellar components and satellites. These predictions will undoubtedly bring new empirical challenges beyond those envisioned here. An operational SIM Lite in the coming decade will be poised as a primary tool for testing these predictions.

## 4.2 Probing the Shape, Profiles, and Lumpiness of Dark Matter Halos

### 4.2.1 Background and Current Problems

Over the last decade, cosmological observations have determined conclusively that normal matter (e.g., protons, neutrons, atoms, etc.) constitutes just one-fifth of the matter in the Universe, with the remaining 80 percent of the cosmic matter budget in the form of DM. Despite the fact that candidate particles for DM have been suggested, the identity of this DM is a complete mystery, but its properties affect how galaxies form and how the Universe evolves. Its existence also is among the strongest pieces of evidence that the Standard Model of particle physics is fundamentally incomplete. In this sense, the nature of DM remains one of the deepest problems in both astronomy and particle physics.

The microscopic nature of DM affects the way it clusters around galaxies, and therefore can be probed by astronomical observations. Dynamical measurements with SIM Lite will provide crucial and unique insights into the properties of the DM in and around our Milky Way, with particular sensitivity to how the DM is distributed on galaxy scales. In the leading model of structure formation, the DM today is mostly "cold," and consists of particles that are much more massive than the proton and that can cluster strongly on small scales. CDM models of structure formation predict that the Milky Way should be surrounded by a diffuse extended "DM halo," which itself should be teeming with smaller, self-bound DM clusters called "sub-halos" (Figure 4-2). These sub-halos orbit the Galaxy like satellites and the mass function of the satellite halos is predicted to rise as the mass decreases (as $1/M$, Diemand et al. 2008). The minimum mass and overall count of dark satellites is linked closely to the particle nature of DM itself. For example, if the DM is cold and consists, e.g., of weakly interacting massive particles (WIMPs), then the mass function may rise steadily to a minimum mass that is close to an Earth-mass (Diemand,

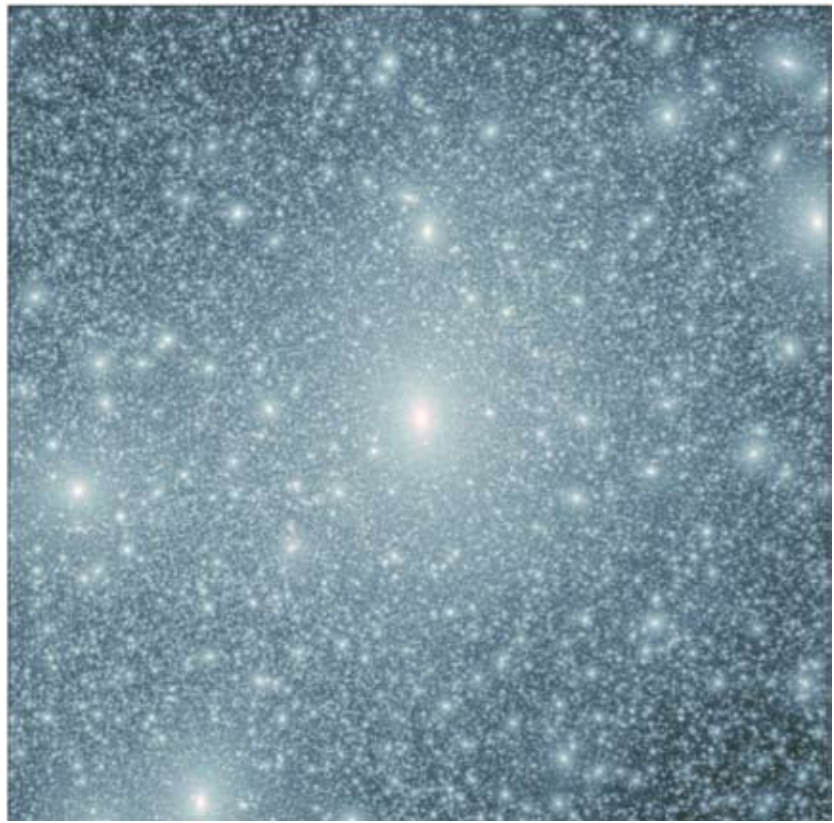

Figure 4-2. The density of DM within the virial radius of a galaxy mass CDM halo simulated with over a billion particles. Over 100,000 substructure halos are present, some orbiting within the inner kpc of the halo. (From Stadel et al. 2008)

Moore, and Stadel 2005), with a total Milky Way count of $N_{sat} \sim 10^{13}$ sub-halos (e.g., Figure 4-2). If, on the other hand, the DM is warm and consists of, e.g., sterile neutrinos, then the mass function may truncate at a mass-scale of $\sim 10^8 \, M_\odot$, with ≤100 self-bound satellite systems in the Milky Way.

SIM Lite observations will provide key insights towards interpreting the so-called "missing satellites problem," which describes the fact that many fewer satellite galaxies are observed around systems like the Milky Way and M31 than the number of CDM sub-halos predicted (Kauffmann, White, and Guideroni 1993; Klypin et al. 1999; Moore et al. 1999; Strigari et al. 2007b). One popular idea is that these dark satellites exist around the Milky Way, but that they are not lit up by stars because star formation is suppressed in the smallest halos (Bullock, Kravtsov, and Weinberg 2000; Benson et al. 2002; Hayashi et al. 2003; Kravtsov, Gnedin, and Klypin 2004). If the majority of sub-halos are dark, then they may be detectable today through their influence on cold Galactic tidal streams (§4.2.2). Moreover, DM particle properties may be determined by SIM Lite measurements of the properties of the satellites that are visible (§4.5).

Our understanding of the satellite galaxy population has been revolutionized in recent years thanks to an increasingly deep census of satellite galaxies around the Milky Way and M31 (e.g., Willman et al. 2005; Martin et al. 2006; Zucker et al. 2006; Belokurov et al. 2007; Majewski et al. 2007). Indeed, the number of known Local Group satellite galaxies has more than doubled since 2005. It remains to be determined how these new galaxies are to be interpreted in the context of the missing satellites problem, and the origin of galaxies as extreme as the ultrafaint dwarfs (e.g., Boo II with $L \sim 1000 \, L_\odot$; Walsh et al. 2008) represents a fundamental challenge to galaxy formation models. It is tempting to evoke tidal interactions to explain their extreme nature; however, these objects are so small that, at their current distance from the Milky Way, they should be relatively unaffected by the Milky Way's tidal force. Only by obtaining proper motion determinations of their orbits with SIM Lite observations (§4.3) can we understand the degree to which these extreme objects were shaped by past central encounters with the Milky Way.

SIM Lite observations will also provide unprecedented constraints on the total mass, mass profile, and shape of the DM distribution around the Milky Way all the way out to its virial radius. Such a detailed, accurate, and extensive assessment of DM distribution is unfeasible for any external galaxy. These constraints are essential in the general goal to use the Milky Way as a "Rosetta Stone" in understanding galaxy formation on a cosmic scale. Mass models based on line-of-sight velocities to distant stars and other tracer objects are fundamentally limited by our ignorance of the orbital eccentricities (i.e., anisotropies, see §4.3–4.4). By determining proper motion measurements to distant halo objects, SIM Lite will provide a definitive measure of the mass profile of the Milky Way. This measurement (with proper, calculated error assessments) will allow us confidently to compare our galaxy to other distant galaxies where mass measurements are more straightforward, but where detailed information on stellar populations and other observables are more poorly determined. Moreover, the mass determination may be compared directly with mass profile predictions for DM halos in the concordance CDM model. Once determined, a globally self-consistent Milky Way dark halo model (with proper error bars) will provide information on the mass range and microphysical properties of the DM particle, and thereby be essential for experiments like Cryogenic Dark Matter Search (CDMS13)[*], XENON14[†], and Large Underground Xenon (LUX15)[‡] that aim to detect the DM particle directly in laboratories on Earth.

In addition to mass profiles, CDM predicts that Milky Way–sized DM halos are typically triaxial, becoming rounder at larger radii (Allgood et al. 2006; Macciò, Dutton, and van den Bosch 2008). Current constraints on the shape of the Milky Way DM halo from modeling the debris of the Sagittarius dwarf spheroidal galaxy are hampered by incomplete phase space information (i.e., we have radial velocities but no proper motions) and vary depending on how the extant data are analyzed, yielding shapes from prolate (Helmi 2004), to near-spherical (Fellhauer et al. 2006), to oblate (Law, Johnston, and Majewski 2005; Martínez-Delgado et al. 2004). SIM Lite proper motions of stars throughout the entire Sagittarius stream, stars in other (more distant) tidal streams (§4.2.2), and hypervelocity stars (§4.2.3) will provide a definitive measurement of the halo shape, orientation, and mass profile with radius. This measurement will not only test model predictions, but will determine how the shape of the inner Galactic halo is influenced by dissipation and the formation of the Galactic disk (Kazantzidis et al. 2004; Zentner et al. 2005, §4.3).

---

*[*] http://cdms.berkeley.edu/     [†] http://xenon.astro.columbia.edu/     [‡] http://lux.brown.edu/*

### 4.2.2 Solution with SIM Lite: Probing the Galactic Potential with Tidal Streams

It is now well known that, as predicted by models of structure formation (e.g., Figure 4-3), the luminous Galactic halo is inhomogeneous and coursed by streams of debris tidally pulled out of accreted satellites. SIM Lite will give us the first opportunity to measure the proper motions of stars in entire streams with very high accuracy (Gaia can measure only portions of the very closest streams accurately). This will allow us to probe two different aspects of the Milky Way's potential: (1) its global shape, which drives the overall phase space distribution of extended tidal streams, and (2) the lumpiness of the potential, which imprints its signal in the degree of small-scale incoherence among member stars at each point in the stream.

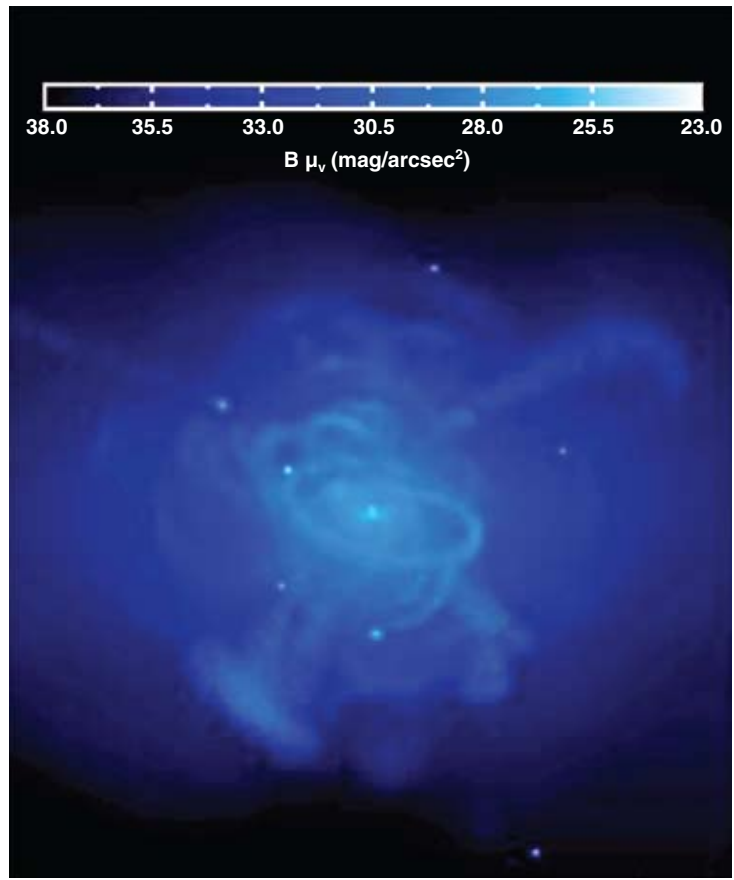

Figure 4-3. Surface brightness map of a simulated stellar halo formed within a ΛCDM context (Bullock and Johnston 2005). The image is 300 kpc on each side, and shows just the stellar halo component (total luminosity ~$10^9$ $L_\odot$). Note that (compared to Figures 4-2 and 4-7), the phase space distribution of stars around the Galaxy is predicted to contain more clear streams than the DM distribution in part because stars form deeply embedded within DM halos. The outer parts of these halos are stripped to form a smooth background component, while the inner parts are able to maintain their coherence longer. In addition, only a small subset of sub-halos are expected to contain any stars at all. (Image credit: Sanjib Sharma, Columbia University)

Once precision proper motion measurements from SIM Lite are combined with the stars' observed angular positions, radial velocities, and estimated distances (accurately calibrated by SIM Lite), stream stars provide a uniquely sensitive probe of the global distribution of mass in the Milky Way. Starting from its full phase space position, each star's orbit can be integrated backwards in some assumed Galactic potential. Only in the correct potential will the stream stars ever coincide in position and velocity with the parent satellite and each other (see Figure 4-4). Tests of this idea with simulated data observed with the µas/yr precision proper motions possible with SIM Lite and km/s radial velocities suggest that 1 percent accuracies on Galactic parameters (such as the flattening of the gravitational potential and the circular speed at the Solar Circle) can be achieved with tidal tail samples as small as 100 stars (Johnston et al. 1999).

Applying this method to several streams at a variety of Galactocentric distances and orientations with respect to the Galactic disk would allow us to build a comprehensive map of the distribution of DM in the halo — its shape, density profile, and orientation. In the last few years evidence for dSph galaxy tidal tails has been discovered around Milky Way satellites at very large Galactic radii (Muñoz et al. 2006; Muñoz, Majewski, and Johnston 2008; Sohn et al. 2007), and other distant streams are known (e.g., Newberg et al. 2003; Pakzad et al. 2004; Clewley et al. 2005). With SIM Lite, such distant streams can be used to trace the Galactic mass distribution as far out as the virial radius with an unprecedented level of detail and accuracy. This would provide the very first, accurate three-dimensional observational assessment of the shape and extent of a galactic-scale DM halo.

Stars in tidal tails also promise to place limits on the possible existence of a significant fraction of the halo in the form of dark satellites (Figure 4-2; Moore et al. 1999; Klypin et al. 1999). Large, dark lumps in the Milky Way's potential should scatter stream stars and lessen their local coherence (Ibata et al. 2002; Johnston, Spergel, and Haydn 2002; Mayer et al. 2002). For this experiment, the coldest streams (e.g., from globular clusters like Palomar 5: Grillmair and Dionatos 2006) would be most sensitive. Early tests of such scattering using only radial velocities of the Sagittarius stream suggest a Milky Way halo smoother than predicted (Majewski et al. 2004), but this result is based on debris from a satellite with an already sizable intrinsic velocity dispersion and incomplete knowledge of the full space motions of the stars, so the constraint is not strong.

Gaia will allow the various tests described here to be attempted for some nearby portions of streams, but only with SIM Lite will it be possible to probe the three-dimensional shape, density profile, extent of, and substructure within our entire galaxy's DM halo.

### 4.2.3 SIM Lite Solution: Probing Galaxy Potentials with Hypervelocity Stars

A complementary method for sensing the shape of the Galactic potential can come from SIM Lite observations of hypervelocity stars (HVSs; Gnedin et al. 2005). Hills (1988) postulated that such stars would be ejected at speeds exceeding 1000 km/s after the disruption of a close binary star system deep in the potential well of a massive black hole, but HVSs can also be produced by the interaction of a single star with a binary black hole (Yu and Tremaine 2003). Recently, Brown et al. (2006) report on five stars with Galactocentric velocities of 550 to 720 km/s. They argued persuasively that these extreme-velocity stars can only be explained by dynamical ejection associated with a massive black hole, and they make similar arguments for three additional stars in Brown et al. (2007). After the success of these initial surveys to find HVSs, it is likely that many more will be discovered in the next few years. If these stars indeed come from the Galactic center (Figure 4-5), the orbits are tightly constrained by knowing their point of origin. In this case the nonspherical shape of the Galactic potential — due in part to the flattened disk and in part to the triaxial dark halo — will induce nonradial inflections (which will be primarily in the transverse direction at large radii) in the velocities of the HVSs of order 5 to 10 km/s, which corre-sponds to of order 10 to 100 µas/yr depending on the HVS distance (Gnedin et al. 2005). Each HVS thus provides an independent constraint on the potential, as well as on the solar circular speed and distance from the Galactic center (Figure 4-6). The magnitudes of the known HVSs range from 16 to 20, so their proper motions will be measurable by SIM Lite with an accuracy of a few µas/yr, which should define the orientation of their velocity vectors to better than 1 percent. With a precision of 20 µas/yr, the orientation of the triaxial halo could be well-constrained and at 10 µas/yr the axial ratios will be well-constrained (Gnedin et al. 2005).

Figure 4-4. A demonstration of the use of tidal streams to measure the shape and strength of the Galactic potential when precise six-dimensional information is available for stream stars. First, a Sgr-like tidal stream was created by the disruption of a dwarf satellite in a rigid Galactic potential through an *N*-body simulation (e.g., as in Law et al. 2005), shown at the top of each strip. In reality, the potential is unknown, but guesses of the strength and shape of the Galactic potential can be tested against the complete, six-dimensional phase space information in hand for the stream stars (including, crucially, SIM Lite proper motion measures). The orbits of the individual stars in the tidal streams can be run backwards under these assumed potentials. The left-hand strip demonstrates what happens if the strength of the Galactic potential is overestimated: when the orbit is run backwards, the tidal stream stars orbit at too large a radius and do not converge (in the bottom panel) to a common phase space position. The right-hand strip shows the result when a Galactic potential of the correct strength is guessed: when the stream star orbits are run backwards, the tidal stream stars collect back into the core of the parent satellite (in the bottom panel).

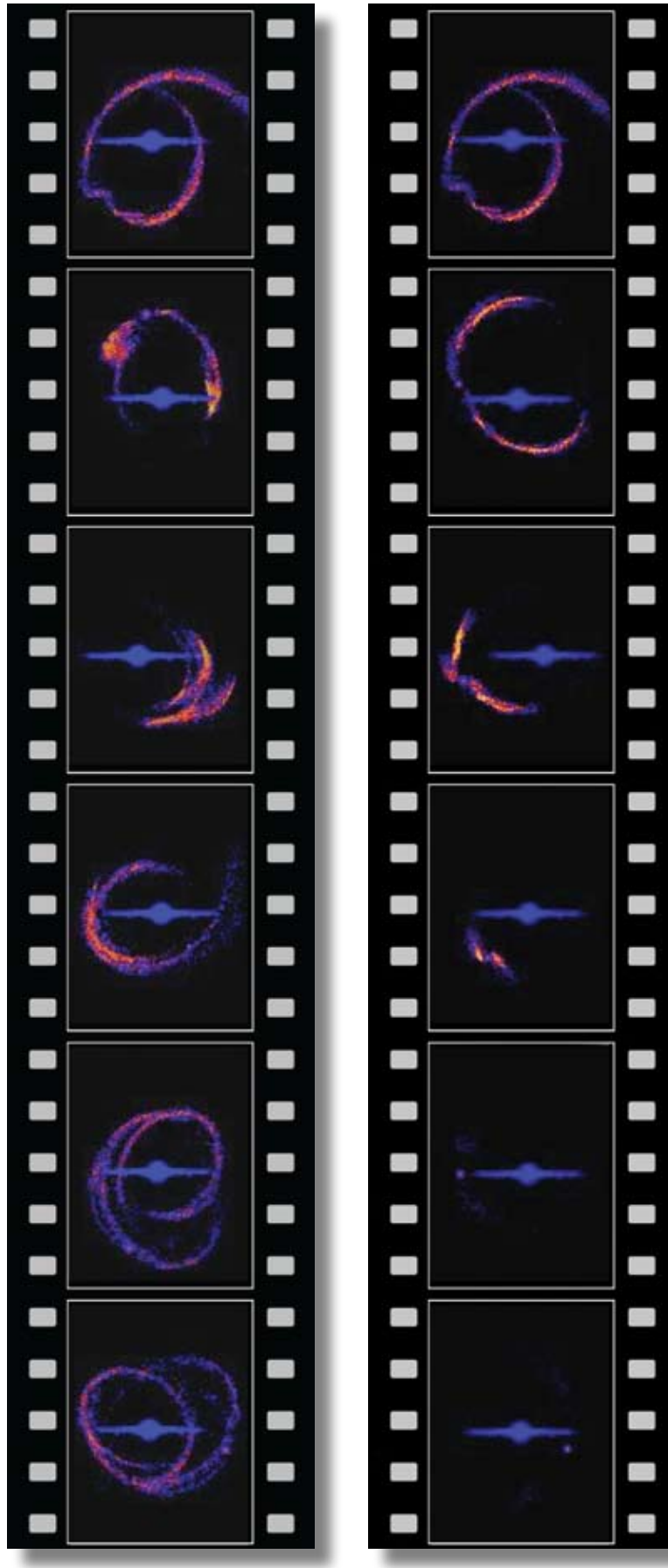

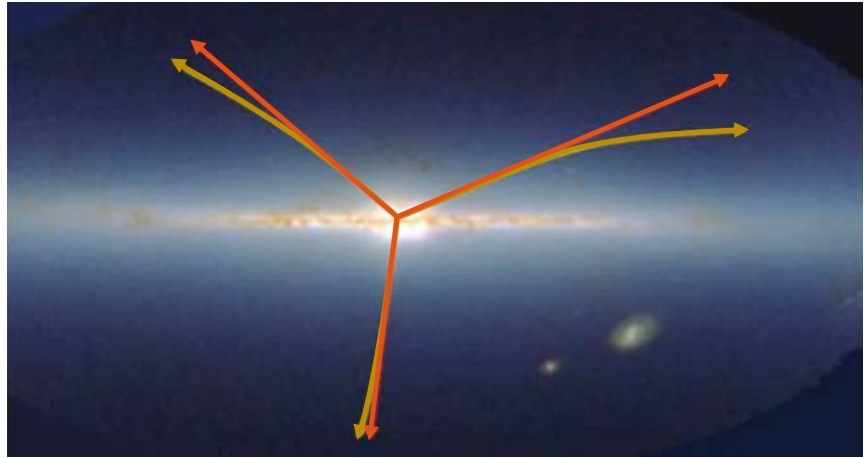

Figure 4-5. Schematic representation of the trajectories of hypervelocity stars in a spherical (red) and non-spherical, flattened (brown) potential. The three-dimensional positions and velocities of hypervelocity stars will, with the inclusion of precise proper motions measured by SIM Lite, have great sensitivity to the global shape of the Galactic potential (see Figure 4-6).

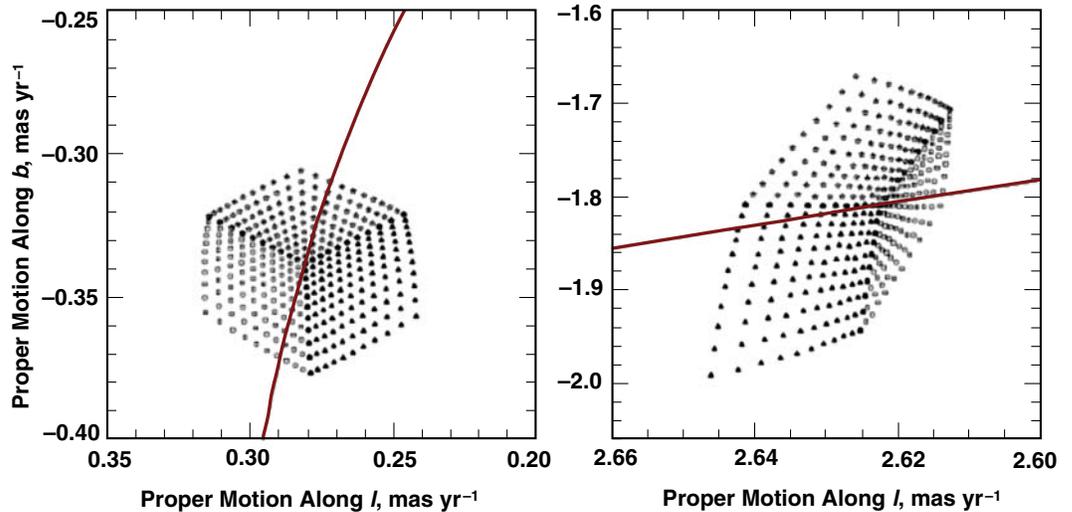

Figure 4-6. Expected proper motions of two known hypervelocity stars, HVS1 (left) and HVS2 (right), under a range of different assumptions (shown as the different symbols) about the triaxial shape and orientation of the Galactic potential, and for one of a range of possible HVS distances (represented by the red lines; see Gnedin et al. 2005). SIM Lite measurements of proper motions can readily select from the families of models shown here.

More recently the existence of an HVS from the LMC has been reported (Gualandris and Portegies Zwart 2007; Bonanos et al. 2008), and the existence of numerous HVSs from M31, including thousands within the virialized halo of the Milky Way, has been suggested (Sherwin, Loeb, and O'Leary 2008). Thus, the next decade will probably bring the identification of many more HVSs of Galactic and extragalactic origin. For the LMC, establishing the three-dimensional trajectory of its HVSs will allow identification of the location of any massive black hole(s) in this galaxy, either in the galaxy center or in the centers of its star clusters. An accurate understanding of the velocities of hypervelocity stars may also lead to an improved understanding of the black hole content of galactic nuclei — e.g., theoretical models show that HVSs will have a different spectrum of ejection velocities and ejection rates from a binary black hole versus a single massive black hole.

## 4.3 Testing Hierarchical Formation and Late Infall

### 4.3.1 Background and Current Problems

With the latest high-resolution *N*-body CDM simulations, it has been possible to make rather detailed predictions for how hierarchical formation of the DM halos should proceed on galaxy scales (Figure 4-7). The infall of DM onto the Milky Way has left a fingerprint of its accretion history that allows for specific tests of CDM if one associates the satellite galaxies (which typically have high mass-to-light ratios) with CDM sub-halos. In the models, sub-halos preferentially align with and are accreted along the intermediate scale filamentary structure (Knebe et al. 2004).

The satellite galaxies of the Milky Way show a clear spatial anisotropy that has often been interpreted as a signature of the specific infall history of the Milky Way's own DM substructures, but this interpretation also strongly depends on the adopted galaxy evolution models applied within the CDM framework (e.g., Hartwick 2000; Libeskind et al. 2005). Moreover, the currently best-available proper motions of the satellite galaxies hint at a strong correlation between their orbits. This might point to (1) the observed satellites not only having fallen in along filaments, but perhaps having fallen into the Milky Way in a few groups of DM sub-halos (Li and Helmi 2008; D'Onghia and Lake 2008) — much as has long been debated for the Magellanic Clouds — or (2) perhaps very different explanations not driven by CDM considerations where alignments are the result of the break-up of formerly larger satellites or the formation of "tidal dwarf" galaxies (Kunkel 1979; Lynden-Bell 1983; Metz, Kroupa, and Libeskind 2008).

Clearly, to understand and test either of these competing theories of the origin of the satellite galaxies, a good understanding of the orbital properties of galaxies in and near the Milky Way is needed. We can establish from their orbits whether and how these satellites correspond to the predicted hierarchically infalling dark sub-halos and what particular infall scenario dominated the formation of the Milky Way's DM halo, or establish whether the dSph galaxies were formed through some other mechanism, e.g., in an early tidal interaction with another galaxy. Orbits for systems at several 100 kpc can also define a complete census of those systems bound to the Milky Way, which places limits on the mass and mass profile of the Galaxy and may be able to probe the hypothesis that there is a third large galaxy in the Local Group hidden behind the bulge of the Milky Way (Loeb and Narayan 2008).

The notion that halo globular clusters were formed in separate environments and were then accreted by the Milky Way over an extended period of time has been a central thesis of Galactic structure studies since the work by Searle and Zinn (1978). More recently, at least some clusters have been identified as part of the Sagittarius stream and the Monoceros structure — which are apparently relatively recent mergers having readily identifiable debris streams — but presumably such mergers have been ongoing and were even more common in the early Galaxy. At present, $\leq 25$ percent of Galactic globular clusters have had any attempt at a measured proper motion, and reliable data generally exist only for those clusters closest to the Sun (see, e.g., Palma, Majewski, and Johnston 2002). With full orbital information for the globular clusters, we can learn not only about their origin, but about the earliest stages of the formation of our own galaxy as well.

While orbit determinations will be useful for ascertaining the origins of globular clusters and dwarf satellites, they are also crucial for understanding the fate of such systems. The dynamical evolution of small stellar systems is largely determined by external influences such as tides and dynamical shocks from the disk and bulge (e.g., Gnedin, Lee, and Ostriker 1999), so determining orbits of these satellites will dramatically improve our understanding of their evolution and address the long-standing issue of whether the present population of these systems is the surviving remnant of a much larger initial population. By finally obtaining definitive orbital data for the entire Milky Way globular cluster and dwarf galaxy sample, SIM Lite will make an important contribution to our understanding of the full range of cluster/satellite dynamical histories on the evolution of the Galactic ensemble. Of course, all of these satellite systems can also serve as valuable test particles for constraining the properties of the Galactic gravitational potential (§4.2).

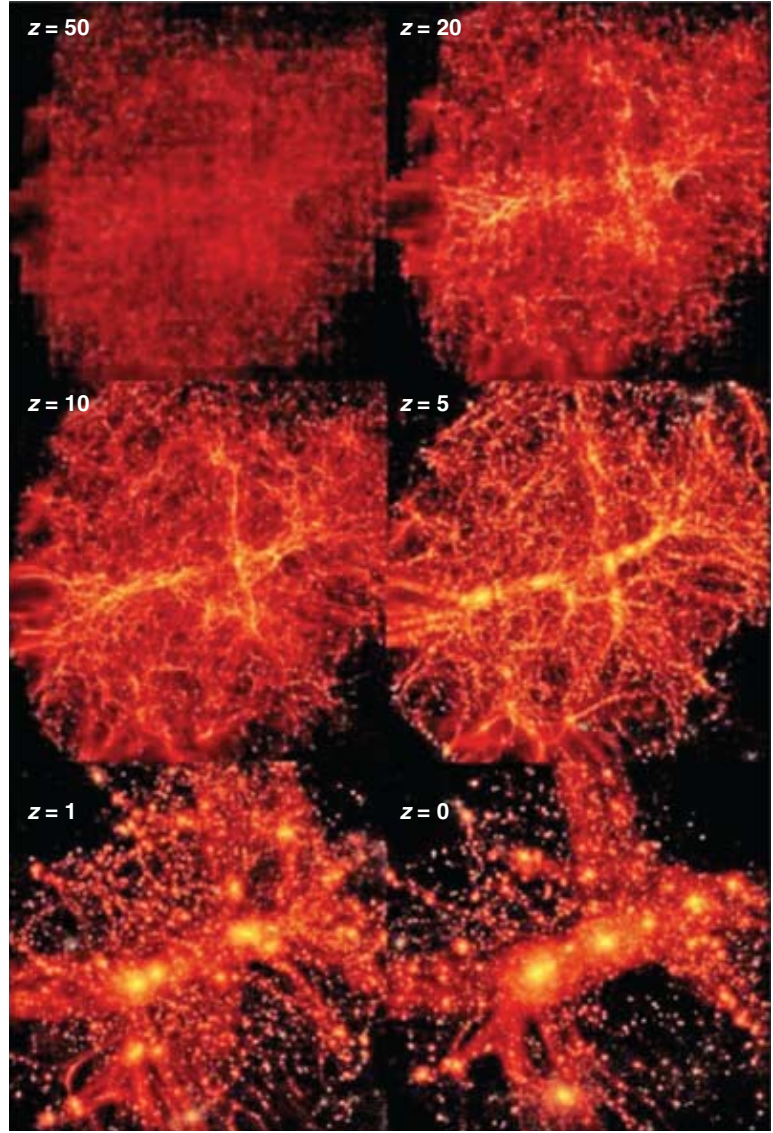

Figure 4-7. The hierarchical formation of a binary galaxy system (showing only the DM) chosen to have similar separation, masses, and infall velocities as the Milky Way and Andromeda galaxies (Moore et al. 2001). The assembly of structures begins at a redshift $z \sim 100$ with, at first, tiny Earth-mass halos, and proceeds until a redshift $z \sim 1$ with the final formation of the main galactic mass halos that are 15 orders of magnitude more massive.

### 4.3.2    Potential Solutions with SIM Lite: Legacy Measurements of Satellite Orbits

The proper motions of the more distant satellites of the Milky Way are expected to be on the order of ~100 µas/yr. To derive transverse velocities good to ~10 km/s (~10 percent) requires a bulk proper motion accuracy of ~10 µas/yr for the most distant satellites at several hundred kpc distances (e.g., Leo I, Leo II, Canes Venatici, Leo T), in which the brightest giant stars have V ~ 19.5. These are systems that require SIM Lite for an accurate proper motion measurement. The constraints are more relaxed for satellites at roughly 100 kpc distances, ~20 µas/yr, with brightest stars at V ~ 17.5. Such measurements are well within the capabilities of SIM Lite, which could derive the desired bulk motions for Galactic satellites with only a handful of stars per system, and, with more stars, be sensitive to systemic rota-tion and other internal motions (§4.4). With its µas-level extragalactic astrometric reference tie-in, and the ability to derive proper motions at the level desired for Galactic satellites with each star measured, SIM Lite will make definitive measures of the proper motions of the Galactic satellites that will overcome complications faced by previous satellite proper motion studies. Moreover, most of the outer halo globular clusters, the numerous recently detected lower surface brightness dSph satellite galaxies (e.g., the Boötes and Hercules dwarfs), and some of the "transitional" type systems (which may be globular clusters or dwarf galaxies — e.g., Willman 1, Coma Berenices) have very sparse giant branches, which prohibits the opportunity to obtain reliable bulk motions by averaging the motions of numerous members (e.g., with Gaia). Only an instrument like SIM Lite, which delivers extremely precise per star proper motions to V ~ 20, can take on this challenge.

## 4.4 Testing the Distribution of Angular Momentum and Anisotropy in Galaxies

### 4.4.1 Background and Current Problems

A series of papers over the past decade by many different theoretical groups has now led to a converging scenario for understanding the origin and substructure of galactic halos within the context of our hierarchical Universe (e.g., Moore et al. 2006, and references within). This understanding is the result of coupling high-resolution numerical simulations with semi-analytical models for star formation within the first massive DM halos capable of hosting global star and cluster formation and dwarf galaxies. Using this technique we can make predictions for the kinematical and spatial distribution of the stellar remnants of these early star-forming structures in galaxy halos today.

We can begin to understand their spatial distribution and kinematics in a hierarchical formation scenario by associating the protogalactic fragments envisaged by Searle and Zinn (1978) with the rare peaks able to cool gas in the CDM clumps collapsing at redshift $z > 10$. The global effects of reionization at this epoch have been suggested to strongly affect the future evolution of the baryons, preventing them from cooling within subsequent generations of dark halos until a much larger mass scale is attained, equivalent to that of the Magellanic Clouds. As a result, the stellar components and globular clusters of our protogalaxy are distributed amongst the ~100 protogalaxies that inhabit these rare peaks of the density field. The thousands of similar mass halos that collapse after this time remain completely dark since the baryonic fluid is too hot to cool efficiently. This scenario could explain the missing satellites problem, i.e., why the bulk of the many thousands of DM substructures visible in Figure 4-2 are not observed.

Most of the rare, luminous protogalaxies rapidly merge together, their stellar contents and DM becoming smoothly distributed and forming the Galactic stellar and dark halo (Figure 4-8). The metal-poor globular clusters and old halo stars formed in these early Milky Way structures become tracers of this early evolutionary phase, centrally concentrated and naturally reproducing the observed steep number density fall off with radius. The most outlying substructures fall in late and survive to the present day as our familiar satellite galaxies. The observed radial velocity dispersion profile and the local radial velocity anisotropy of Milky Way halo stars are successfully reproduced in this model, but only with full three-dimensional orbits can we be assured that the orbital shapes are truly consistent with predictions.
If this epoch of structure formation coincides with the suppression of gas cooling into halos that collapse after reionization, then we can reproduce the rarity, kinematics, and spatial distribution of satellite galaxies, as suggested by Bullock et al. (2000). Reionization, therefore, provides a natural solution to the missing satellites problem and the origin of the stellar halo and its old globular cluster population.

The angular momentum distribution of the merging and infalling substructures also plays an important role in determining the structure and sizes of galactic disks and by this, in turn, galaxy morphology. Cur-rently, numerical cosmological simulations of disk galaxy formation encounter a serious problem with too much angular momentum being transferred from the visible to the dark component during infall, leading at the end to galactic disks that are too compact and not in agreement with observations (e.g., Abadi et al. 2003). In order to find a solution for this problem it is crucial to determine the average specific angular momentum and the mass distribution of angular momentum in galactic halos.

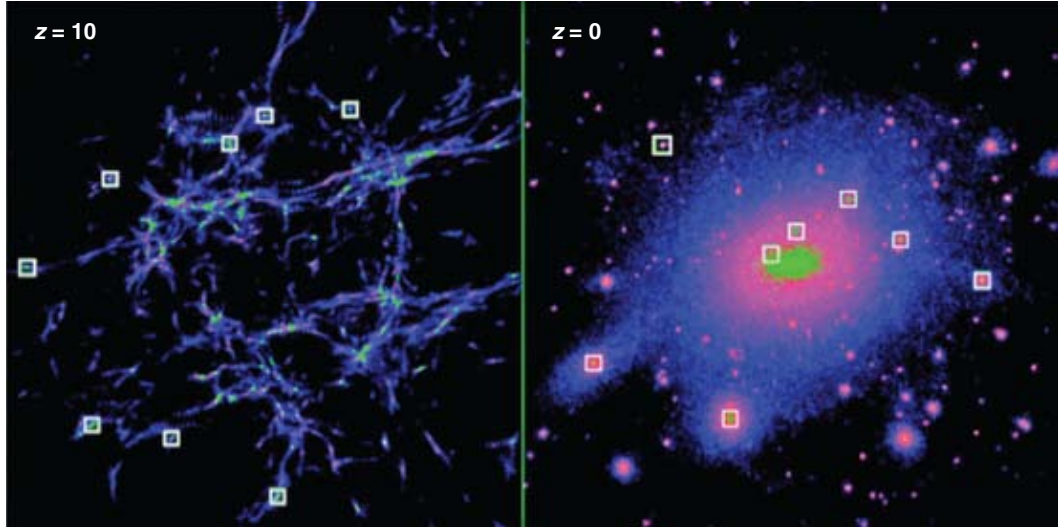

Figure 4-8. The evolution of dark and luminous sub-halos in a Milky Way–like galaxy. The halos shaded in green in the left panel are those that could have cooled gas and formed stars before reionization at $z = 10$. The right panel shows the location of the particles in those halos at $z = 0$ — most of them have merged together, but the boxes show the luminous satellites that survive the merging process. They were all the most distant from the forming halo at $z = 10$ and survive simply because they fell in late, after virialization. The ~100 other protogalaxies at $z = 10$ shaded in green merged together, with their debris (globular clusters and stars) spread out to form the $\sim r^{-3}$ density law halo (most visible in the right panel by its highest-density central region). Each panel is 250 kpc on a side. (From Moore et al. 2006)

### 4.4.2 Potential Solutions with SIM Lite: The Angular Momentum Profile of the Milky Way

Not only can SIM Lite test the predictions of the model for early galaxy formation described above, but it can also disentangle the stellar orbits in phase space in order to reconstruct the accretion history and properties of these protogalaxies. The expected properties of the spatial distribution and kinematics of material within today's Galactic halo that originated within these first CDM protogalaxies that formed before reionization are precisely predicted from high-resolution numerical simulations such as Diemand et al. (2005). According to these models, the outermost Galactic halo stars and globular clusters are on radially biased orbits, while the inner regions contain more isotropic orbits. The kinematics of the predicted surviving satellite dwarf galaxies that fell in late is different, with small anisotropies even in the outer parts. In addition, the net angular momentum of these components is expected to be small and simply reflect that of the dominant DM component at this epoch. These orbital differences between the early and late infalling structures are a simple test for SIM Lite.

A relatively small sample of proper motions for halo stars and the halo globular clusters would check and constrain further these theoretical models. The angular momentum and anisotropy measured within a sample of a few hundred proper motions to a precision of 5 km/s (compared to the radial dispersion of ~100 km/s) in different regions of the sky and at different Galactic radii would allow us to model accurate orbits, obtain orbital angular momenta and eccentricities, and provide a detailed statistical accounting of the initial specific angular momentum of sub-halos. SIM Lite can provide these critical data on the angular momentum and anisotropy at large radii, well beyond the Gaia-sphere. Moreover, SIM Lite may be able to provide a similar assessment of the independent angular momentum profile of the Magellanic Clouds, providing insight into the structure and formation of intermediate mass halos.

## 4.5 Measuring the Physical Nature of Dark Matter

### 4.5.1 Background and Current Problems

Determining the nature of the DM that surrounds galaxies and overwhelms the gravitational force from visible matter constitutes a fundamental task of modern astrophysics. CDM particles are characterized by low initial velocity dispersion and high phase space density, resulting from a relatively heavy particle mass. After collapsing, they thermalize into distributions with steep density cusps at their centers (Navarro et al. 2004; Diemand et al. 2005). In cosmologies with a somewhat lighter DM particle there are reduced phase space densities and higher velocity dispersions. These alternative models, broadly classified as warm DM, produce more constant density cores in galactic halos (Tremaine and Gunn 1979).

In this sense, by precisely measuring the shape of the central DM density profile (characterized by the logarithm of the slope there), one places important constraints on the primordial phase space density of DM, which in turn bears on such microphysical properties of the DM particle as its mass and details of its formation mechanism. For example, in the case where the DM was in thermal equilibrium with the plasma in the early Universe (this would be the case with "thermal warm DM"), then the limit comes down to a constraint on the particle mass. If the DM particle is born in a decay, then the phase space constraint limits a combination of the decay lifetime and the particle mass difference between the daughter and the primary (as in the case of SuperWIMPS, where the gravitino is the DM and it emerges as a daughter particle from the late decay of a heavier particle).

Based on measured mass-to-light ratios, dwarf spheroidal (dSph) galaxies occupy the least massive known DM halos in the Universe. Dwarf spheroidals are also unique among all classes of galaxies in their ability to probe the particle nature of DM, because phase space cores resulting from the properties of the DM particle are expected to be most prominent in these small halos (Tremaine and Gunn 1979; Hogan and Dalcanton 2000). In recent years, the measurement of line-of-sight velocities for upwards of a thousand stars in several dSphs has allowed for a precise determination of their masses (e.g., Strigari, Bullock, and Kaplinghat 2007a; Walker et al. 2007). However, despite the great progress made in estimating masses of these systems, determining the logarithmic slope of their central density profile, and thus the nature of the DM contained within, remains elusive.

Modeling of the mass distribution of the dSphs requires a solution of the equilibrium Jeans equation, which gives a relation between the components of the intrinsic velocity dispersion and the mass density profile of the DM halo. When projecting along the line of sight, the observed velocity dispersion is a combination of two orthogonal components, and thus a degeneracy arises between the ratio of the radial and tangential components of the velocity dispersion and the logarithmic slope of the DM density profile (Figures 4-9, 4-10, 4-11). Even with 1000 line-of-sight velocities, the relative error on the log-slope at the King core radius is of order $\approx 1$, which cannot distinguish between a cusp model, which has typical log-slope of $\approx 1$ at this radius, and a core model, which has a typical log-slope of $\approx 0.3$ at this radius (Figure 4-10, upper panels; Strigari et al. 2007a).

### 4.5.2 Potential Solutions with SIM Lite: Dark Matter Within Dwarf Galaxies

The only way to break this degeneracy is to sample the tangential velocity components in addition to the line-of-sight components. Even were the numbers of stellar radial velocities available in individual dSphs eventually to reach of order $10^5$, there would be insufficient numbers at and within the critical radius (about 100 pc) that must be probed to discriminate reliably a central core from a central cusp. The only way to break the degeneracy between the log-slope of the density profile and the velocity anisotropy of dSph stars is to sample the tangential velocity components in addition to the line-of-sight components. A measurement along these lines requires the proper motion of individual stars to a precision of at least 10 km/s, which is the intrinsic velocity dispersion of the highest luminosity dSphs.

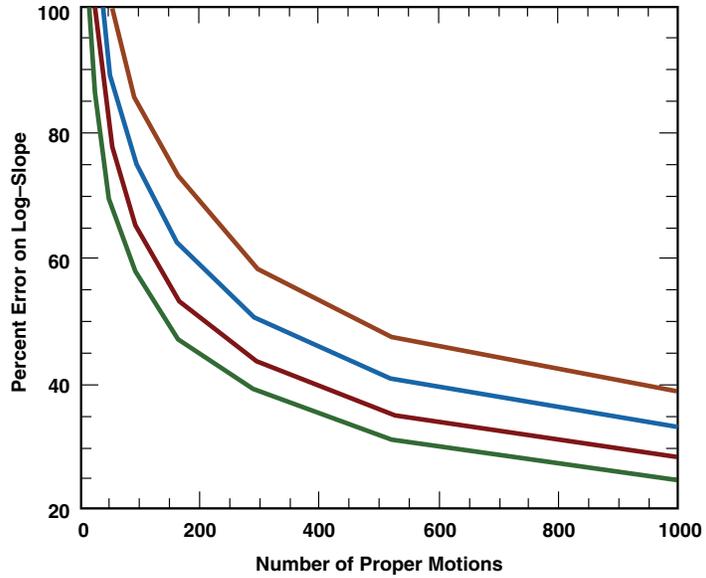

Figure 4-9. The projected error on the log-slope of the DM density profile of Draco at the King core radius as a function of the number of proper motions. From top to bottom, the curves assumed that the errors on the transverse velocities are 10, 7, 5, 3 km/s.

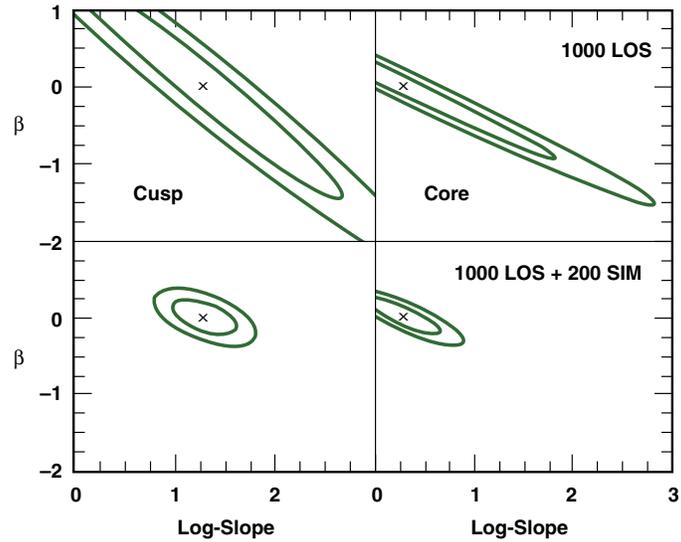

Figure 4-10. A demonstration of the ability to recover information on the nature of DM using observations of dSph stars, from analytical modeling by Strigari et al. (2007a). Ellipses indicate the 68 percent and 95 percent confidence regions for the errors in the measured dark halo density profile slope (measured at twice the King core radius) and velocity anisotropy parameter $\beta$ in the case where only radial velocities are available for 1000 stars in a particular dSph (top panels). A significant improvement is derived from the addition of 200 SIM Lite proper motions providing 5 km/s precision transverse velocities (bottom panels). The left (right) panels correspond to a cusp (core) halo model for dSphs and the small ×'s indicate the fiducially input model values.

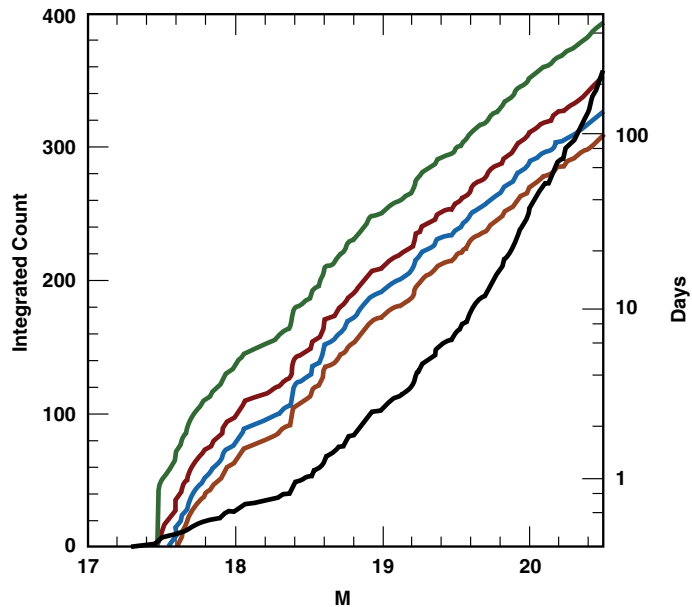

Figure 4-11. Potential SIM Lite exploration of the Draco dSph would need to probe to V ~ M = 19 to derive a sample of 200 red giants as seen by its Washington M-band luminosity function (black line and left axis). The colored lines represent the number of days (right axis) necessary to observe all the stars to a given magnitude limit with SIM Lite and for a given transverse velocity uncertainty: 3 km/s (green line), 5 km/s (red line), 7 km/s (blue line) and 10 km/s (magenta line). (From Strigari et al., in preparation.)

Several dSphs within 100 kpc, in particular Draco (Figure 4-11), Sculptor, Carina, and Ursa Minor, have hundreds of stars with V < 20, and thus proper motions of these stars may be obtained with SIM Lite. SIM Lite will be able to measure the proper motions of hundreds of stars in several of these dSphs to a precision of ~5 km/s. In Figure 4-9, we show the projected error on the log-slope of the DM density profile at the King core radius as a function of the number of measured proper motions for a typical dSph. The number of proper motions is in addition to 1000 line-of-sight velocities. As may be seen, with several hundred proper motions from one dSph it will be possible to measure the log-slope to a precision of ~30 percent, which is several times better than a measurement of the log-slope with line-of-sight velocities alone (Strigari et al. 2007a). Thus, with the investment of 1500 hours (Figure 4-11), SIM Lite should be able to discriminate cusped from cored density profiles in the Draco dSph halo (Figure 4-10). Obtaining the required transverse velocities, while well beyond the capabilities of Gaia, is well-matched to the projected performance of SIM Lite.

## 4.6  Dark Matter and Dynamics on the Galaxy Group Scale

### 4.6.1  Background and Current Problems

Within the emitting parts of galaxies, rotation curves are generally flat rather than falling according to Kepler's law, which implies that mass grows roughly linearly with radius. The total mass-to-light ratio depends critically on where this mass growth ends, but this is generally not observable. Just beyond the outermost measured 21 cm isophotes of galaxies, the DM distribution simply becomes unknown. As a result, critical questions about DM on scales from galaxy to galaxy-group sizes remain unanswered. Do dwarf galaxies have lower or higher mass-to-light ratios than regular galaxies? Do the DM halos of galaxies in groups merge into a common envelope? How do these mass components compare with the warmer DM particles smoothly distributed across super-clusters or larger scales?

At present, we can only detect DM through its gravitational effects; therefore, a careful study of the dynamics of nearby galaxies is one of the few ways to resolve these issues. It is reasonable to anticipate that numerical simulations (e.g., constrained *N*-body and least-action method calculations) will find the initial density distribution that transforms into a best fit to present-day redshifts and proper motions, once they are known, for nearby individual galaxies and groups of galaxies. This fit will yield masses for the galaxies and galaxy groups, and will also give the density of uniformly distributed warm/hot DM at Mpc scales. Although the hot DM density is thought to be quite small today, its gravitational signature from early times may be significant because the density of relativistic material grows as $(1 + z)^4$, and, in a high $\Omega_\Lambda$ universe, the cross-over from energy-dominated to matter-dominated occurs at a later time.

Most analyses of peculiar velocity flows have applied linear perturbation theory appropriate for scales large enough that over-densities are << 1 (Peebles 1980). Peculiar velocity analyses (Shaya, Tully, and Pierce 1992; Dekel et al. 1993; Pike and Hudson 2005) have proven the general concept that the observed velocity fields of galaxies result from the summed gravitational accelerations of over-densities over the age of the Universe. Peculiar velocity analyses agree with virial analyses of clusters and Wilkinson Microwave Anisotropy Probe (WMAP) observations that indicate the existence of a substantial DM component strewn roughly where the galaxies are. But these studies apply only to scales of >10 Mpc, the scales of superclusters and large voids. Spherical infall (including "timing analysis" and "turnaround radius") studies do not presume low over-densities and have been applied to the smaller scale of the Local Group (Lynden-Bell and Lin 1977). These studies indicate mass-to-light ratios for the Local Group of roughly $M/L \sim 100\ M_\odot / L_\odot$ in blue light, but this model is crude; nonradial motions and additional accelerations from tidal fields and subclumping are expected to be non-negligible and would substantially alter the deduced mass (Li and Helmi 2008). A more complete treatment of solving for self-consistent orbits is required, but to properly constrain such models observations of nonradial motions is a prerequisite.

### 4.6.2  Potential Solutions with SIM Lite: Measuring Galaxy Motions Within the Local Group

If one measured the proper motions of nearby Local Group galaxies with global accuracies of a few µas/yr, one would have another pair of high-quality phase space components with which to construct flow models and determine histories and masses for galaxies and galaxy groups. For a galaxy 1 Mpc away, 4 µas/yr corresponds to 19 km/s transverse motion, which is small compared to the expected transverse motions in the field, ~100 km/s. Therefore, SIM Lite will be able to measure accurate proper motions of any galaxy in the Local Group that has stars brighter than V < 20. The low-velocity dispersion of stars in a typical dwarf galaxy ensure that after averaging just a few random stars, the contribution to the error from the internal motions would be only a few km/s. For larger galaxies, simple rotation models adjusted to the observed velocity profiles, or direct measurements of the rotation with SIM Lite (§6.4), can be removed from the motions to deliver <20 km/s accuracy. There are 27 galaxies beyond the Milky Way satellites (all within 5 Mpc) that have sufficiently bright stars.

SIM Lite measurements of the three-dimensional galaxy flows will be of lasting importance in the modeling of the formation of the Local Group, several nearby groups, and of the plane of the Local Supercluster. Gaia will probably be able to obtain proper motion vectors for satellite galaxies of the Milky Way, M31, M33, and WLM by averaging a sufficient number of stars in these systems that are bright enough. For other galaxies, we would expect Gaia to obtain upper limits based on a few dozen stars near its detection limit. The proper motion measurement of M31 by either SIM or Gaia would be limited essentially by errors in the referenced coordinate grid. It is expected that the SIM frame would be a factor of >2 more accurate than Gaia's. The Gaia measurement should prove to be a significant improvement over the VLBI study of two water masers in M33 (Brunthaler et al. 2005).SIM Lite's ability to point for long periods at targets will allow it to obtain high-accuracy proper motions for the other galaxies, which have just a few stars with V < 20.

The application of the least-action method (LA, Peebles 1989) allows one to solve for the trajectories that result in the present distribution of galaxies (or more correctly, the centers of mass of the material that is presently in galaxies). By making use of the constraint that early-time peculiar velocities were small, the problem becomes a mixed boundary value problem with constraints at both early and late times. Figure 4-12 shows the output of a recent LA calculation for the orbits of nearby galaxies and groups going out to the distance of the Virgo Cluster. The orbits are in co-moving coordinates. This is just a single example of a set of several solutions using present three-dimensional positions as inputs. The masses of four massive objects (Virgo Cluster, Coma Group, CenA Group, and M31) have been adjusted to provide a best fit to observed redshifts. The good news is that complex orbits are rare. With the addition of proper motions of 27 galaxies, the problem becomes fully constrained, and one can solve for the actual masses of all of the dominant galaxies. Experiments have shown that, with 5 percent distance accuracy, errors in mass are below 50 percent for the dominant galaxies. In other words, we will finally learn where the DM ends on each major galaxy. Also, the two components of proper motion will be useful for determining the mass associated with groups aside from that in the individual galaxies, as well as the amount of matter distributed on scales larger than 5 Mpc.

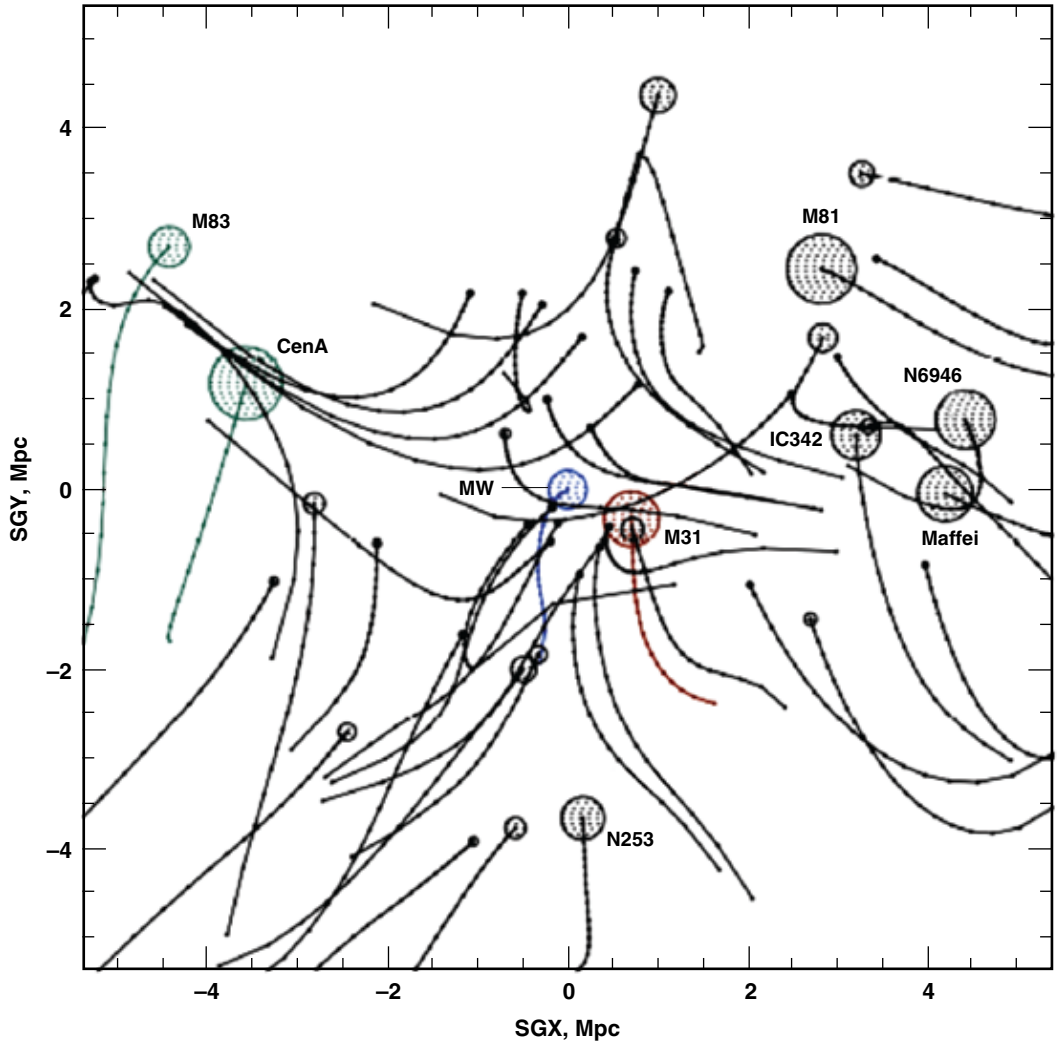

Figure 4-12. The trajectories of nearby galaxies and groups going out to the distance of the Virgo Cluster from a Least Action calculation with parameters $M/L = 90$ for spirals, 155 for ellipticals, $\Omega_m = 0.24$, and $\Omega_\Lambda = 0.76$. The axes are in the supergalactic plane (SGX-SGY) in co-moving coordinates. There are 21 time steps going from $z = 40$ to the present. The large circle is placed at the present position and the radius is proportional to the square root of the mass. The present estimated distances were fixed and the present redshifts were unconstrained.

## 4.7   Summary of Proposed SIM Lite Possibilities and Priorities

While the discussion above has outlined a number of important problems within the study of Local Group dynamics and DM, it may not be possible to address them all with a fixed mission time. Table 4-1 summarizes the experiments discussed in this chapter and the estimated SIM Lite mission times required to approach each problem in a definitive and legacy-creating way. It must be kept in mind that as this active field evolves, the relative priorities of these different experiments — and indeed the consideration and priority of additional experiments as yet uncontemplated — cannot be anticipated for a decade hence. Thus, we do not attempt to prioritize these proposed tests here, but only tabulate the expected mission requirements for each. Whatever sub-suite of these will be selected at final mission definition, SIM Lite will make unique, lasting, and significant contributions to the study of local DM and galaxy formation.

As Table 4-1 summarizes, the ability of SIM Lite to make µas measurements of the positions of faint stars is unmatched among all planned missions. SIM Lite alone will be able to probe accurately the six-dimensional stellar phase space distribution around the outskirts of our own Galaxy, within a nearby sample of dwarf galaxies, and in other Local Group systems — and these are the only galaxies in the Universe for which such data will be available. These measurements provide unique windows on some key astrophysical problems, from galaxy formation and the structure of DM halos, to the nature of DM itself.

Table 4-1. Estimated SIM Lite observing time for galactic dynamics and local DM experiments.

| Experiment | No. of Stars | Magnitude Range, V | PM Accuracy, µas/yr | Mission Time, hrs |
|---|---|---|---|---|
| Tidal Streams (§4.2.2) | 400 | 11–19 | 4–20 | 700 |
| Milky Way High-Velocity Stars (§4.2.3) | 20 | 18–20 | 10–60 | 300 |
| Extragalactic High-Velocity Stars (§4.2.3) | 20 | 18–20 | 10–60 | 300 |
| Satellite Orbits (§4.3.2) | 350 | 14–20 | 7–40 | 600 |
| Angular Momentum Profile (§4.4.2) | 100 | 16–20 | 50–75 | 100 |
| Dark Matter in Dwarf Galaxies (§4.5.2) | 200 | 17–19 | 11 | 1500 |
| Local Group Galaxy Motions (§4.6.2) | 600 | 17–20 | 2–10 | 1000 |